# Fault Tree Synthesis from Knowledge Graphs

Manzi Aimé Ntagengerwa, University of Twente, the Netherlands

Georgiana Caltais, University of Twente, the Netherlands

Mariëlle Stoelinga, University of Twente and Radboud University, the Netherlands

## SUMMARY & CONCLUSIONS

A truly effective diagnostic system provides system engineers with valuable insights into the behavior of their machines, leveraging a rich body of (often tacit) expertise. Much of this expertise typically resides in written documentation or troubleshooting manuals, which are frequently imprecise or vaguely specified. Therefore, methods for formalizing this knowledge, such as through the use of knowledge graphs, are of particular interest. However, ensuring that the extracted knowledge (ideally in a semi-automatic way) encapsulates sufficient semantic depth for system-level diagnostics is a challenging task. In this paper, we propose a minimal format for knowledge graphs that is semantically rich enough to facilitate the synthesis of meaningful fault trees. Fault trees offer an intuitive and efficient means for systematic failure analysis, enabling engineers to assess all potential failure modes in a structured, hierarchical manner. The methodology is applied to the Lycoming O-320 engine, showing that meaningful fault trees can be synthesized from only structural and functional knowledge of the system, defined by the proposed conceptual model.

## 1 INTRODUCTION

Cyber-Physical Systems (CPSs) are complex machinery that integrate computational and physical functionality [1]. Companies commonly design such machines from a compositional and functional perspective. For instance, Model-Based Systems Engineering (MBSE) is a methodology that supports engineering activities through the use of connected models [2]. SysML [3][4] and Arcadia [5] are popular examples of MBSE languages that allow for the structural and functional decomposition of a system. Tooling allows for validating properties of the model, such as consistency with respect to model constraints.

The task of structurally and exhaustively modeling explicit failure mechanisms is time-consuming and challenging, often crossing boundaries between engineering teams and domains of expertise. As a result, many organizations do not have sufficient explicit knowledge about failures at a system level. Failure Mode and Effects Analysis (FMEA) [6] is a common method of failure analysis. The result of applying this method is a document with text and tables (called worksheets), which require manual upkeep throughout a system lifecycle. While FMEA allows for the structural analysis of failure modes, worksheet entries are large and difficult to read [7]. The construction of FMEA documents is a time-consuming and tedious task [8] and depends largely on effective communication an alignment between the different teams and domain experts in the organization. It is error-prone and many potential failure mechanisms are not uncovered [9]. Since FMEAs are often created and maintained in plain text, they may suffer from inconsistencies and contradictions. Copy-paste errors are common, as are ambiguities and misunderstood semantics between different authors of FMEAs [7]. The lack of explicit fault knowledge of the system under design (especially across domain boundaries) can lead to missed opportunities in designing a robust system.

The integration of system design knowledge from different sources provides additional ways of exploring failure mechanisms at a system level. Knowledge graphs (KGs) are a popular method of integrating such knowledge across design teams and domain boundaries. However, visualizations of KGs are typically large and hard to interpret. (Formal) reasoning over a knowledge graph is possible, but developing the right reasoning rules is a challenging and time-consuming task. An effective model for explaining possible failure modes and mechanisms is the Fault Tree (FT). The semantics of FTs are well defined, with many formal analysis methods available out-of-the-box [10]. However, manually creating FTs is time-consuming, hard and error-prone.

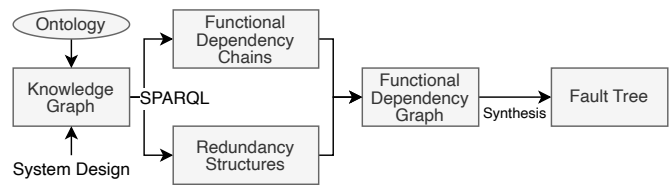

*Figure 1 – The methodology of transforming KGs to FTs*

*Our approach.* We propose a methodology to automatically transform knowledge graphs into fault trees, see Figure 1. Our methodology starts with a knowledge graph, presenting the structural decomposition and functional specification of CPSs. The ontology of the KG describes exactly how a CPS's components, functions, and resources interact, and captures enough knowledge to synthesize meaningful FTs. In particular, the KG allows us to automatically discover redundancy in the system. Technically, failure propagation is discovered through SPARQL queries. These extract data from the KG used to build two models: a graph of functional dependency chains and a graph of redundancy structures. Both are then combined to synthesize fault trees. This automatic transformation greatly reduces the effort involved in obtaining and maintaining FTs.

*Running Example.* We will use the (simplified) design of the Lycoming O-320 aircraft engine as a running example to illustrate our methodology. The system knowledge and data formats available for the Lycoming engine is representative of the aforementioned issues, where there are no models of the system under design, only plain-text descriptions and diagrams [11][12], and knowledge of the different aspects of the engine (domains) is not integrated in one model. The ignition subsystem is tasked with igniting the fuel mixture in the cylinder to create a mechanical force. The engine is designed with the goal of resiliency to failure in mind, and is equipped with two redundant sets of ignition systems. They operate in parallel and largely independent from each other. This particular engine is equipped with magnetos, which convert the cylinder's mechanical energy into electrical energy. This electricity is then transported to the spark plugs, which in turn generate a spark to ignite the fuel mixture. We demonstrate that our proposed methodology yields meaningful FTs, generated from the structural and functional decomposition of the Lycoming O-320 aircraft engine.

*Overview of the paper.* We provide related work in Section 2. We introduce background on knowledge graphs and fault trees in Section 3 and Section 4 respectively. Section 5 introduces our proposed ontology and relates CPS design knowledge to failure propagation. Section 6 shows our method for transforming knowledge graphs to fault trees.

## 2 RELATED WORK

Several approaches exist to obtain fault trees from system models. The work in [13] describes a method to synthesize FTs from SysML's Internal Block Diagrams (IBD). IBDs represent a system's internal structure and the interactions between its components through ports. This approach relies on specific SysML semantics, while we argue that the conceptual model proposed in our work is generally applicable to KGs that contain structural and functional system knowledge, regardless of the modeling language in which such system knowledge is originally expressed. This allows us to integrate different models (in different modeling languages) that describe the same system. The work in [14] allows for the automatic generation of FTs from SLIM models (an extension of AADL). While these FTs are expressive and accurate, the synthesis methodology requires detailed error models of a system. In contrast, our approach infers FTs from a static system model.

Hip-HOPS [15] aims to address the inconsistency of failure analysis techniques that are often found in practice. The methodology integrates Functional Failure Analysis (FFA), Failure Mode and Effect Analysis (FMEA), and Fault Tree Analysis (FTA). Hip-HOPS introduces a novel algorithm for fault tree synthesis. However, the Hip-HOPS methodology involves manual analysis of system models and requires domain knowledge to uncover explicit fault knowledge.

When fault semantics are encoded in an ontology, like in [16], generating FTs from KGs that instantiate such an ontology may be more straight-forward. The ontology presented in [17] explicitly encodes FT concepts. However, both these ontologies restrict the applicability to KGs where explicit fault knowledge is already present. The synthesis of FTs from KGs based on only structural and functional system designs is a key value proposition of our approach.

## 3 KNOWLEDGE GRAPHS

A knowledge graph (KG) is a knowledge-based system that integrates information into an ontology and allows for the derivation of new knowledge using a reasoner [18]. A well- known example of a KG is Wikidata [19], the underlying technology on which Wikipedia is built. An ontology is a conceptual model that describes how concepts in a certain domain relate to each other. A reasoner is a system of logical inference that makes explicit facts from implied knowledge. For instance, a KG containing the facts *(John, isA, Person)* and *(Julia, isA, Person)*, together with the inference rule *all persons are equal* leads to the addition of a new, explicit fact in the KG *(John, isEqualTo, Julia)*. KGs typically comprise three layers of abstraction.

- The *ontology layer* is the highest level of abstraction, where real-world concepts and the relationships between them are described.
- The second layer is called the *taxonomy*, and provides a hierarchy of types. Together with the ontological layer, this forms conceptual model.
- The lowest abstraction level contains instances of the conceptual model.

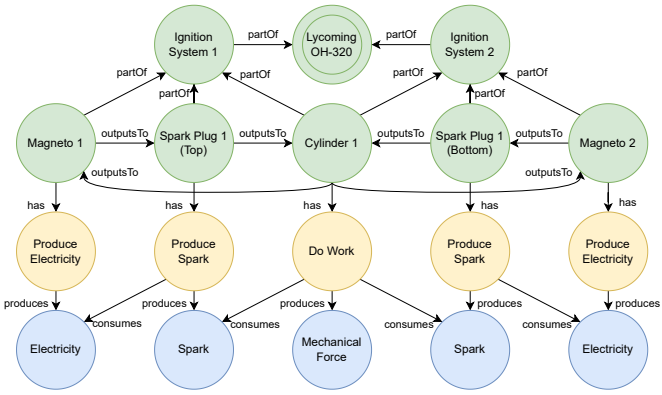

*Figure 2 – a KG of the running example*

*Example.* Figure 2 shows a visualization of the KG describing (a part of) the Lycoming O-320 engine. The figure shows a hierarchy of components (modeled by the *partOf* relation), the exchange of resources (determined by the *inputFrom* and *outputsTo* relations), and functions and resources (modeled by the *has*, and the *produces* and *consumes* relations respectively). The predicate edge label corresponds to a relationship described in the ontological layer of the KG. For instance, the nodes *"Ignition System 1"* and *"Lycoming O-320"* in Figure 3, related through the *partOf* relation are encoded in an RDF graph as the triple *("Ignition System 1", partOf, "Lycoming O-320")*. The coloring of the nodes in Figure 2 is a simplified visual representation of the actual RDF graph. Figure 3 shows the meaning of these colors. For example, the green color of the node *"Magneto 1"* is shorthand for the existence of an additional RDF fact *("Magneto 1", rdf:type, Component)* that is not shown.

## 4 FAULT TREES

A Fault Tree (FT) is a deductive, hierarchical model used to analyze the causes of system-level failures. It represents the logical relationships between component failures and overall system failures using a graphical structure. The root node of a fault tree typically denotes a critical failure event (Top Event), and the leaf nodes represent basic component failures or events. Logical gates, such as AND and OR, are used to connect these events, illustrating how multiple failures can combine to result in higher-level failures [10]. For OR gates, the output event is observed if any of its input events occur. For AND gates, the output event is only observed if all of its input events occur. Fault trees are extensively used in safety and reliability engineering to systematically analyze potential faults. By mapping failure pathways, they help identify vulnerable components, assess risk probabilities, and propose mitigation strategies. Fault Tree Analysis (FTA) is a prominent technique in risk analysis [10]. FTA methodologies are applied in many engineering and governmental domains [20][21]. FTs can facilitate qualitative analysis by, for instance, helping to determine minimal cut sets – the smallest combinations of basic events that can cause the top event. FTs could also support quantitative analysis, such as calculating the probability of the top level event occurring based on the probabilities of basic events. Such probabilistic analysis methods only work if statistical fault data of basic events is available.

In our study, we synthesize Fault Trees inferred from the structural and functional system knowledge encoded in a KG. To keep the knowledge assumptions minimal, we assume no statistical knowledge of fault events, making the synthesized FTs particularly suitable for qualitative analysis.

## 5 ONTOLOGY FOR THE FUNCTIONAL DESIGN OF CPS

An important ingredient in our framework is the ontology, being the first layer of the knowledge graph, see Figure 3. Our proposed ontology describes the structural composition of components in CPSs and associates them with their expressed function. Our ontology is tailored to capture sufficient knowledge of CPS failure propagation and generate meaningful FTs. It is common in industrial practice to model systems in a compositional and functional way [22].

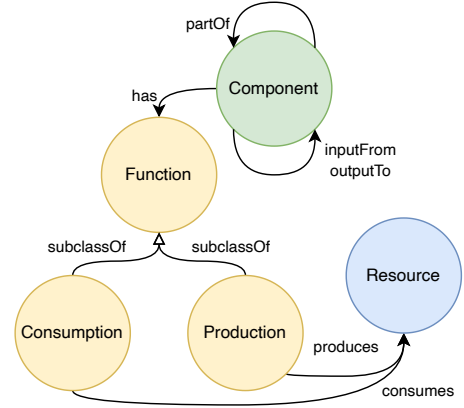

*Figure 3 – The proposed ontology*

*Components* are digital or physical objects that have the ability to express some function. Components often comprise other components as constituents. A component without constituents is called a *part* or *atomic* [23]. As such, a part is the boundary of where a designer stops modeling finer structural details of a component. It is up to a designer to determine what the boundaries of a component are. We call a component that is not a constituent of any other component a *system*.

A *function* describes the ability of a component to deliver a service. That service is delivered by transforming some input resource into some output resource. We specify the types of input and output resources, analogous to how we specify a mathematical function's domain and range.

A *resource* is any form of energy, such as material, electricity, or information. This is a broad definition that encompasses many of the substances, data and interactions involved in the operation of a CPS. Examples of concrete resources include *gas*, *water*, *pressure*, *control commands* or *data streams*. In the modeling process, we sometimes do not consider a function's undesired or inconsequential inputs and outputs. For instance, we may not model the waste heat that bicycle brake pads produce. Where to draw such system boundaries is a design choice.

We distinguish two special types of functions, denoted by the *subclassOf* relation: *consumption* and *production*. If a function consumes some resource, it is a consumption. If a function produces some resource, it is a production. A concrete function can be an instance of both a consuming and a producing function. In the running example, all functions are both consuming and producing.

The *partOf* relation describes the composition of components. For any two components $c_1$ and $c_2$, we say that $c_1$ is a constituent of $c_2$ if and only if *($c_1$, partOf, $c_2$)*. This relation is transitive, such that for any components $c_1$, $c_2$, $c_3$, it holds that *($c_1$, partOf, $c_2$) ∧ ($c_2$, partOf, $c_3$) →  ($c_1$, partOf, $c_3$)*. As a result, $c_1$ is a constituent of both $c_2$ and $c_3$.

The *has* relation denotes the ability of a component to express a certain function. In this work, we assume that if a component has the ability to express a function, it does express the function. Similarly, if a function has the ability to consume or produce some resource, it does consume or produce that resource. Therefore, in the rest of this paper the has relation may be interpreted as "the transformation of resources continuously performed by a component". We call these assumptions liveliness. They enable the generation of FTs from a static view of the system.

The *consumes* and *produces* relations indicate the particular resources that a function consumes and produces, respectively. A function can consume and produce any number of different resources.

The *inputFrom* and *outputsTo* relations indicate the ability of components to exchange resources. This resource exchange is constrained by the ability of one component to supply a resource (production), and the ability of another component to accept that resource (consumption). Under the liveliness assumption, all functions, including production and consumption functions, are expressed continuously, so we extend the liveliness assumption to include this resource exchange: if components have the ability to exchange resources, they do exchange resources.

### 5.1 Failure Terminology

This subsection relates the standard notions of *faults*, *errors*, and *failures* to the concepts of *functional dependency* and *failure propagation*. Following [23], a *failure* is an event that occurs when a system's delivered service deviates from its correct service. An *error* is a part of the system state that may lead to a system failure. The cause of an error is called a *fault*. Faults that do not cause an error are called *dormant*.

However, since under the liveliness assumptions all functions are constantly enabled and expressed, a fault will always immediately cause an error. In other words, there are no dormant faults. Because a fault will, under these assumptions, always induce an error, we combine the two concepts into one for convenience: a *fault event* is defined as the (spontaneous) inability of a component to express its designed functions.

During operation, a fault event may occur at a component, resulting in a component failure. We define *failure propagation* as the effect we observe when a component failure is in itself an event that leads to the failure of another component. This is similar to the definition of external error propagation in [23].

Fault events correspond to *Basic Events* (BE), *intermediate events* and *Top-Level Events* (TE) in FTs. We call a fault event a BE when it occurs at a component whose functions are not implemented through delegation to a constituent component. We call a fault event an intermediate event when it is the logical result of the occurrence of one or more other fault events. We call a fault event a TE when it is an intermediate event that occurs at a component that is a system.

### 5.2 Functional Dependency and Failure Propagation

For a component $c_2$ that has a consumption function, in order to consume some resource $r$, that resource needs to be available to $c_2$. Resources are made available through a resource exchange between components, described by the two IO relations. For this resource exchange to happen, there must be a component $c_1$ that has a production function $p$ that produces the resource $r$. We say that component $c_2$ is functionally dependent on $c_1$ for resource $r$, denoted $c_1 <_r c_2$.

The functional dependency between two components gives rise to a causal failure relationship between components $c_1$ and $c_2$. We can now define failure propagation under the liveliness assumption: for any two functionally dependent components $c_1 <_r c_2$, when a fault event occurs at component $c_1$, preventing it from producing the resource $r$ that $c_2$ depends on, there are two scenarios:

1. There exists no other component on which $c_2$ is dependent that produces resource $r$, and $c_2$ fails, or;
2. There does exist another component $c_3$ on which $c_2$ is dependent that produces the resource $r$, and therefore $c_2$ continues to function.

The question of whether a failure propagates from $c_1$ to $c_2$ is answered by determining the existence of a third component $c_3$, such that we can confirm or refute the statement $c_3 <_r c_2$.

*Chains of Functional Dependency*: A sequence of components with a pair-wise functional dependency forms a chain of functional dependency. Assume that each component in this chain receives their required resources from only a single component. By the definition of failure propagation, when any component in this chain fails, all subsequent components in that chain can no longer express their designed functions and also fail. This aligns with the fault semantics of the FT's OR gate, where the fault propagates to a higher level of the tree when any one of its child nodes fails.

*Functional Redundancy*: A component can be function- ally dependent on more than one component for the same resource. In this paper, failure propagates to $c_n$ if and only if all components that supply resource $r$ to $c_n$ fail. This is equivalent to a conjunction of the failures of all components $c_k$ where $c_k <_r c_n$. This matches the semantics of the FT's AND gate: faults propagate to a higher level of the tree when all of its child nodes fail. Note that this may over-approximate the robustness of the system.

## 6 METHODOLOGY

The transformation of a KG to an FT takes three steps. First, we query the KG to extract a functional dependency graph from it. This enables the construction of OR gates. Second, we identify redundancy structures in the KG and add them to the dependency graph, enabling the construction of AND gates. Third, we map the nodes in this functional dependency graph to BEs, intermediate events, TEs, and gates. The mapping of functional dependency structures to FT elements follows the definition of failure propagation (Section 5.2). We use SPARQL [24] to query the KG. This semantic query language enables complex graph queries, supporting pattern matching, filtering and aggregation, all founded in formal semantics [25]. A query can be written based on the KG's ontology, similar to how schemas define structure in relational databases. We use queries to extract functional dependency chains and redundancy structures from the KG. One can observe a straight- forward correspondence between the query triples and the definition of failure and failure propagation described in Section 5.2.

### 6.1 The Functional Dependency Graph

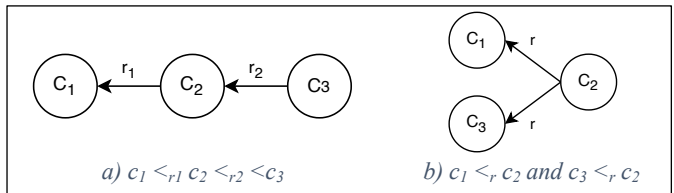

*Figure 4 – Functional dependency chains (a) and redundancy structures (b).*

The functional dependency graph is a simple data model (a Del graph) where each node is a component, and each edge denotes a "depends on" relationship between its tail component and its head component. The label associated with each edge describes the resource(s) for which the tail component is dependent on the head component, as illustrated in Figure 4a. Listing 1 shows the SPARQL query that extracts functional dependency. Applying this to the running example results in Figure 5.

```
1  SELECT ?c1 ?io ?c2 WHERE {
2      ?c1 rdf:type :Component.
3      ?c2 rdf:type :Component.
4      ?c1 :has ?f1.
5      ?f1 rdf:type :Production.
6      ?f1 :produces ?resource.
7      ?c2 :has ?f2.
8      ?f2 rdf:type :Consumption.
9      ?f2 :consumes ?resource.
10     {?c1 ?io ?c2.} UNION {?c2 ?io ?c1.}.
11     FILTER (?io IN (:inputFrom, :outputsTo)) }
```

*Listing 1 – SPARQL query extracting functional dependency*

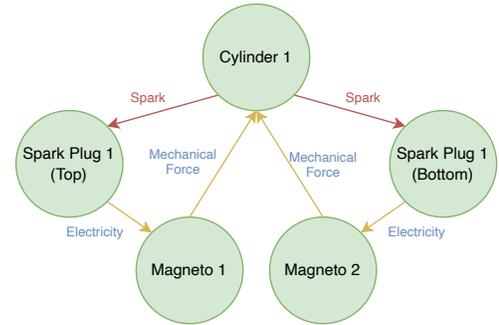

*Figure 5 – Functional dependency graph for the running example. In red, a redundancy structure.*

At this point, the functional dependency graph may implicitly contain redundancy structures. The next step is to identify them in the KG and make them explicit. An example is shown in Figure 4b. We use the query shown in Listing 2 to identify redundancy structures in the functional dependency graph, illustrated by the red edges in Figure 5.

```
1   SELECT ?c1 ?c3 ?resource ?c2 WHERE {
2       ?c1 rdf:type :Component.
3       ?c3 rdf:type :Component.
4       ?c2 rdf:type :Component.
5       {?c1 ?io1 ?c2.} UNION {?c2 ?io1 ?c1.}.
6       {?c3 ?io2 ?c2.} UNION {?c2 ?io2 ?c3.}.
7       ?c1 :has ?f1.
8       ?f1 rdf:type :Production.
9       ?f1 :produces ?resource.
10      ?c3 :has ?f2.
11      ?f2 rdf:type :Production.
12      ?f2 :produces ?resource.
13      ?c2 :has ?f3.
14      ?f3 rdf:type :Consumption.
15      ?f3 :consumes ?resource.
16      FILTER (?io1 IN (:inputFrom, :outputsTo)).
17      FILTER (?io2 IN (:inputFrom, :outputsTo)).
18      FILTER (?c1 != ?c3) }
```

*Listing 2 – SPARQL query extracting redundancy structures*

## 6.2 Fault Tree Synthesis

The last step is generating the FT. A component can spontaneously be subject to a fault event, creating an "internal failure" BE for each component. An OR gate is generated, and the "internal fault" BE is added as a child to this. Recall from Section 5.2 that a component may fail if another component that it functionally depends on fails. Whether or not it depends on the availability of a third component to supply that same resource. For any component $c_2$ in the functional dependency graph, find all outgoing edges. For each of these edges, determine if it is in a redundancy structure with $c_2$ for each of its corresponding resources $r_i$. If so, generate an AND gate between the OR gate labeled $c_2$ and all other components in that redundancy structure for $r_i$. Lastly, generate an additional OR gate, a child of the OR gate labeled $c_2$, for each of the components that $c_2$ is functionally dependent on, but are not in a redundancy structure with $c_2$. Applying this methodology to the example results in an FT (Figure 6).

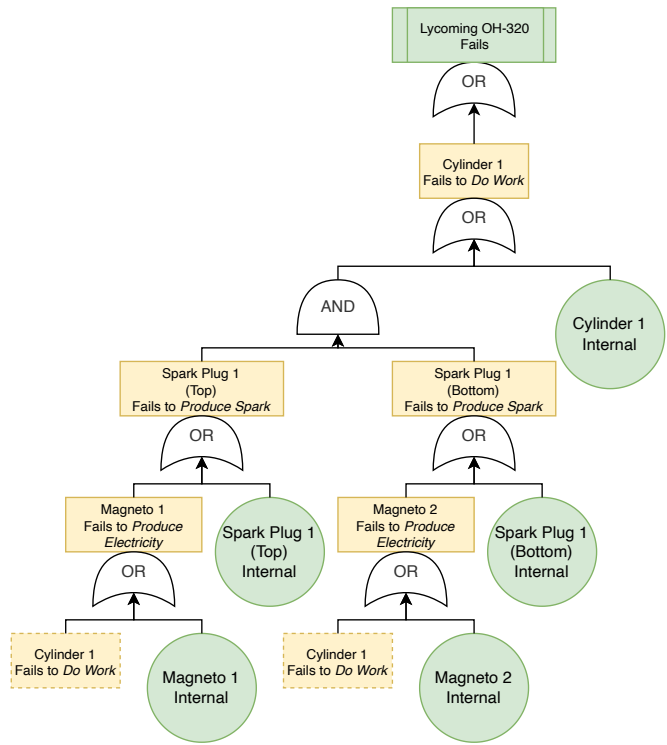

*Figure 6 – Synthesized FT for Lycoming O-320 engine. In yellow, the intermediate events that correspond to component failures. In green, the BEs for internal fault events.*

## 7 CONCLUDING REMARKS

We have shown how meaningful fault trees can be synthesized under a small set of knowledge assumptions. This enables qualitative fault tree analysis methods to be applied over a KG that contains no explicit fault information. The liveliness assumption simplifies this synthesis process. However, the FTs may suggest potential failure modes that do not exist in the actual CPS's operational modes. Future work could investigate whether incorporating system state and time information in the methodology may address this issue, potentially at the cost of requiring more explicit system knowledge in the KG. Furthermore, the inference of redundancy structures may over-approximate the robustness of a system. Our proposed methodology does not distinguish a redundancy structure from a component that requires an identical resource from two distinct sources. Future work may extend the ontology with an explicit redundancy concept. Lastly, we hypothesize that it is possible to match subsets of SysML and Arcadia models to our proposed ontology, making our methodology available to existing system models.

*ACKNOWLEDGEMENT*

This publication is part of the project ZORRO with project number KICH1.ST02.21.003 of the research programme Key Enabling Technologies (KIC) which is (partly) financed by the Dutch Research Council (NWO).

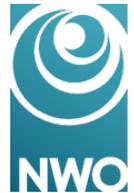